\documentclass[11pt]{article}
\usepackage[margin=1in,left=1in]{geometry}

\usepackage{amsthm}
\usepackage{amssymb}
\usepackage{amsmath}
\usepackage{mathtools}
\usepackage{adjustbox}

\usepackage[table]{xcolor}

\usepackage{stmaryrd}    % for \sslash
\usepackage{rotating}
\usepackage{xfrac}
\usepackage{enumitem}\setlist{nosep}
\usepackage{slashed}

\usepackage[safe]{tipa} % for \esh

\usepackage{pgffor}

\usepackage{hhline}

\usepackage{tensor}

\usepackage{accents}
\newlength{\dhatheight}

\usepackage{tikz}
\usetikzlibrary{
  backgrounds, 
  decorations.pathmorphing, 
  decorations.markings,
  decorations.text,
  snakes
}
\usetikzlibrary{cd}

\usepackage{mathrsfs} %mathscr
\DeclareMathAlphabet{\mathpzc}{OT1}{pzc}{m}{it} %for math script

\usepackage{hyperref}
\hypersetup{
  colorlinks = true,
  linkcolor=darkblue, citecolor=darkblue,
  urlcolor=black
}

\definecolor{darkblue}{rgb}{0.05,0.25,0.65}
\definecolor{darkgreen}{RGB}{20,140,10}
\definecolor{lightgray}{rgb}{0.9,0.9,0.9}
\definecolor{darkorange}{RGB}{200,100,5}
\definecolor{darkyellow}{rgb}{.91,.91,0}

\theoremstyle{definition}

\newcommand{\proofstep}[1]{\scalebox{.8}{#1}}

\newcommand{\HilbertSpace}[1]{\mathcal{#1}}

\newcommand{\ZTwo}{\mathbb{Z}_2}

\newcommand{\rbraid}{
  \begin{scope}[yscale=.5]
  \draw 
  [line width=2pt]
    (-1,1)
    .. controls (.1,1) and (-.1,-1) ..
    (1,-1);

  \draw[white, line width=6pt]
    (-1,-1)
    .. controls (.1,-1) and (-.1,+1) ..
    (1,+1);
  \draw 
  [line width=2pt]
    (-1,-1)
    .. controls (.1,-1) and (-.1,+1) ..
    (1,+1);
  \end{scope}
}
\newcommand{\lbraid}{
  \begin{scope}[yscale=-1]
    \rbraid
  \end{scope}
}
\newcommand{\strand}
{
  \draw 
  [line width=2pt]
    (-1,0) to (1,0);
}

\let\PLAINthebibliography\thebibliography
\renewcommand\thebibliography[1]{
  \PLAINthebibliography{#1}
  \setlength{\parskip}{0.5pt}
  \setlength{\itemsep}{0.5pt plus .3ex}
}

\newcommand{\shape}{
  \raisebox{1pt}{\rm\normalfont\textesh}
}

\newcommand{\defneq}{\equiv}

\newcommand\bosonic[1]{\mathstrut\mkern2.5mu#1\mkern-14mu\raise1.7ex%
  \hbox{$\scriptstyle\rightsquigarrow$}}

\newcommand{\grayunderbrace}[2]{\color{gray}\underbrace{\color{black}#1}_{\color{gray}#2}\color{black}}
\newcommand{\grayoverbrace}[2]{\color{gray}\overbrace{\color{black}#1}^{\color{gray}#2}\color{black}}

\usetikzlibrary{decorations.pathmorphing,shapes}
\newcounter{sarrow}
\newcommand\xrsquigarrow[1]{%
\stepcounter{sarrow}%
\begin{tikzpicture}[decoration=snake]
\node (\thesarrow) {\strut#1};
\draw[->,decorate] (\thesarrow.south west) -- (\thesarrow.south east);
\end{tikzpicture}%
}

\begin{document}
%%%%%%%%%%%%%%%%%%%%%%%%%%%%%%%%%%%%%%%

%%%%%%%%%%%%%%%%%%%%%%%%%%%%%%%%%%%%%%%%%%%%%%
%vertical spacing around displayed equations %
\setlength{\abovedisplayskip}{3pt}
\setlength{\belowdisplayskip}{3pt}
\setlength{\abovedisplayshortskip}{-3pt}
\setlength{\belowdisplayshortskip}{3pt}
%%%%%%%%%%%%%%%%%%%%%%%%%%%%%%%%%%%%%%%%%%%%%%

%%%%%%%%%%%%%%%%%%%%%%%%%%%%%%%%%%%
\title{Topological QBits in Flux-Quantized Super-Gravity}
%%%%%%%%%%%%%%%%%%%%%%%%%%%%%%%%%%%

\author{
  Hisham Sati${}^{\ast \dagger}$,
  \;\;
  Urs Schreiber${}^{\ast}$
}

\maketitle

\begin{abstract}
  We first give a brief exposition of our recent realization of anyonic quantum states on single M5-brane probes in 11D super-gravity backgrounds, by non-perturbative quantization of the topological sector of the self-dual tensor field on the 6D worldvolume, after its proper flux-quantization.
  This opens the prospect of holographic models for topological qbits away from the usual but unrealistic limit of large numbers of branes.

\smallskip 
  At the same time, the elementary homotopy-theoretic nature of the construction yields a slick expression of topological quantum gates in homotopically-typed programming languages, opening the prospect of topological-hardware aware quantum programming.

  \smallskip
  In view of these results, we end with some more meta-physical remarks on (cohesive) homotopy (type) theory in view of emergent fundamental physics and, possibly, M-theory.
\end{abstract}

\vspace{.1cm}
\begin{center}
 invited contribution to:

 \
 
 X. Arsiwalla, H. Elshatlawy and D. Rickles (eds.): 
 
 {\it \color{darkblue}Quantum Gravity and Computation}
 
 Routledge (2025)
\end{center}

\vspace{1cm}

\begin{center}
\begin{minipage}{5.5cm}
\tableofcontents
\end{minipage}
\end{center}

\vfill

\hrule
\vspace{.2cm}

{
\footnotesize
\noindent
\def\arraystretch{1}
\tabcolsep=0pt
\begin{tabular}{ll}
${}^*$\,
&
Mathematics, Division of Science; and
\\
&
Center for Quantum and Topological Systems,
\\
&
NYUAD Research Institute,
\\
&
New York University Abu Dhabi, UAE.  
\end{tabular}
\hfill
\adjustbox{raise=-15pt}{
\href{https://ncatlab.org/nlab/show/Center+for+Quantum+and+Topological+Systems}{
\includegraphics[width=3cm]{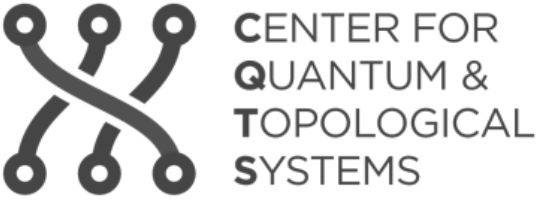}}
}

\vspace{1mm} 
\noindent ${}^\dagger$The Courant Institute for Mathematical Sciences, NYU, NY.

\vspace{.2cm}

\noindent
The authors acknowledge the support by {\it Tamkeen} under the 
{\it NYU Abu Dhabi Research Institute grant} {\tt CG008}.
}

\newpage

\noindent
{\bf An open problem.}
While the hopes associated with {\it quantum computation} \cite{GH19} are hard to overstate, it is a public secret that fundamental new methods are needed for realizing useful quantum computers {\it at scale}. Plausibly, these methods will inevitably need to involve topological stabilization, notably via {\it anyonic} quantum states (e.g. \cite{ZCZW19}\cite{SS23-ToplOrder}, i.e., via solitons whose states pick up purely topological quantum phases when moved around each other). 

\smallskip 
At the same time, despite the resulting attention that the idea of {\bf topological quantum computation} 
\cite{Sau17}\cite{MySS24} (see \cite{SV} for a survey)
has thus received, the microscopic understanding of anyonic topological order has arguably remained sketchy, due to the general lack of first-principles understanding of strongly-coupled/correlated quantum systems -- which may also explain the dearth of experimental realizations of {\it topological q-bits} to date: Better fundamental theory may be needed to understand how anyonic quantum states can actually arise in quantum materials.

%%%%%%%%%%%%%%%%%%%%%%%%%%%%%%%%
\section{Quantum gravity...}
\label{QuantumGravity}
%%%%%%%%%%%%%%%%%%%%%%%%%%%%%%%

\noindent
{\bf M-branes.}
Remarkably, a potential solution --- to this host of theoretical problems arguably impeding practical progress --- has emerged from the study of {\it quantum gravity} (e.g. \cite{West12}). In its locally super-symmetric enhancement, super-gravity (SuGra) shows hints of having a completion to a general theory of strongly-coupled interactions, where the dynamics of strongly-correlated quantum systems may usefully be mapped onto the fluctuations of {\it m}embranes (\cite[\S 2]{Duff99-MTheory}, whence the working title: ``M-theory'' \cite{Duff96}\cite{Duff99-MTheory}) and higher dimensional {\it 5-branes}
\cite[\S 3]{Duff99-MTheory}\cite{GSS24-FluxOnM5}\cite{GSS24-M5Embedding} inside an auxiliary higher dimensional spacetime (11D SuGra \cite[\S 1]{Duff99-MTheory}\cite{GSS24-SuGra}), a phenomenon famous as {\it holographic duality} \cite{ZLSS15}. 

For example, the phase transitions of quantum-critical superconductors, not amenable to traditional weak-coupling (``perturbative'') analysis, have been understood at least qualitatively by these gravitational M-theoretic methods \cite{HKSS07}\cite{GSW10a}\cite{GSW10b}\cite{GPR10}\cite{DGP13}  (review in \cite{Pires14}\cite{ZLSS15}\cite{Nastase17}\cite{HLS18}). More precise quantitative results cannot be expected without an actual formulation of M-theory/holography beyond the usual but unrealistic {\it large-$N$ limit} of a macroscopic number of coincident such branes.

Progress on developing M-theory any further had stagnated, but we may notice that a fundamental non-perturbative phenomenon already in classical super-gravity has received little to no attention in this context, namely the issue of ``flux-quantization''. We find this to be crucial:

\medskip

\noindent
{\bf Flux-quantization.}
While the non-perturbative quantization of gravity famously remains a fundamental open problem of theoretical physics, we may observe that the higher (categorical symmetry) gauge fields that appear in the graviton super-multiplet call for a non-perturbative completion already at the classical level, namely by a {\it flux-quantization} law
\cite{SS25-Flux} 
which determines the topologically stabilized solitonic field configurations.  
This is classical for ordinary electromagnetism (where  {\it Dirac charge quantization} in ordinary cohomology stabilizes the Abrikosov vortices observed in type II super-conductors, cf. \cite[\S 2.1]{SS25-Flux}) and it is famous for the RR-field in 10D supergravity (which a popular conjecture sees flux-quantized in topological K-cohomology, stabilizing certain non-supersymmetric D-branes, cf. \cite[\S 4.1]{SS25-Flux}), but it had received little attention for the C-field in the pivotal case of 11D SuGra (where flux-quantization will stabilize non-supersymmetric M-branes and solitons on M5-brane worldvolumes, cf. \cite[\S 4.2]{SS25-Flux}).

\medskip

\noindent
{\bf Non-abelian cohomology.}
In fact, due to the {\it non-linear} electric Gauss law,
\begin{equation}
  \label{ElectricGaussLaw}
  \mathrm{d}
  \,
  G_7
  \;=\;
  \tfrac{1}{2}
  G_4 \, G_4
  \,,
  \;\;\;\;
  \mbox{where\;
  $
  \mathrm{d}\, G_4
  \;=\;
  0
  \,,\;\;\;\;
  G_7 \;=\; \star G_4
  $}
\end{equation}
on the flux densities sourced by M-branes in 11D super-gravity (cf. \cite[\S 3.1.3]{MiemiecSchnakenburg06}\cite[Thm. 3.1]{GSS24-SuGra}), none of the familiar {\it Whitehead-generalized cohomology} theories may serve here as flux-quantization laws (since they are intrinsically linear, or: abelian); what is needed are instead \cite[\S 3]{SS25-Flux} generalized {\it non-abelian cohomology} theories (\cite[\S 2]{FSS23-Char}, generalizing the ordinary non-abelian cohomology of Chern-Weil theory which classifies gauge  and gravitational instantons)
that are receiving attention only more recently (such as in the study of non-abelian Poincar{\'e} duality \cite{Lurie14}).

\medskip

\noindent
{\bf Classifying spaces and characters.}
The idea behind this powerful concept of non-abelian cohomology becomes quite simple once one realizes that all reasonable cohomology theories have and are characterized by {\it classifying spaces} $\mathcal{A}$ (cf. \cite[p. 19]{SS25-Flux}), so that the cohomology classes on a given space $X$ are just the homotopy classes of maps from $X$ to $\mathcal{A}$, denoted
\begin{equation}
  \label{NonabelianCohomology}
  H^1(X;\, \Omega \mathcal{A})
  \;=\;
  \mathrm{Maps}(
    X
    ,\,
    \mathcal{A}
  )_{\!/ \mathrm{hmtpy}}
  \,.
\end{equation}

A simple but profound rule (cf. \cite[Prop. 3.7]{SS25-Flux},  using the fundamental theorem of dg-algebraic rational homotopy theory, cf. \cite[\S 5]{FSS23-Char}) determines the {\bf admissible flux quantization laws} $H^1(-;\Omega \mathcal{A})$ for given flux densities $(F^{i})_{i \in I}$ satisfying Bianchi identities $\mathrm{d} F^i \,=\,P^i\big(\vec F\big)$: The flux species $F^i$ of degree $\mathrm{deg}_i$ must span the real $\mathrm{deg}_i$-homotopy groups of the classifying space $\mathcal{A}$, and the cohomology of the Bianchi identities on the free graded algebra generated by the flux densities must coincide with the real cohomology of $\mathcal{A}$.

For example (see also \cite[p. 21]{SS25-Flux}), vacuum electromagnetism with $\mathrm{d}\, F_2 \,=\, 0$ requires a classifying space whose real-homotopy and real-cohomology both are generated by a single element in degree 2, such as the universal first Chern class on infinite-projective space $\mathcal{A} \,\defneq\, \mathbb{C}P^\infty \,\simeq\, B\mathrm{U}(1)$ which classifies the ordinary 2-cohomology known from Dirac charge quantization; 
while the NS B-field with $\mathrm{d}\, H_3 \,=\, 0$ may similarly be flux-quantized by the next such {\it Eilenberg-MacLane space} $\mathcal{A} \,\defneq\, B^2 \mathrm{U}(1)$ which classifies ``bundle gerbes''; and the RR-fields with $\mathrm{d} F_{2k} \,=\, H_3\, F_{2k-2}$ require a classifying space with such a generator in every even degree -- such as $\mathrm{KU}_0 \,\defneq\, \underset{\longrightarrow_{\mathrlap{n}}}{\lim} \, B \mathrm{U}(n) \times \mathbb{Z}$ with its higher universal Chern classes -- twisted to incorporate the $H_3$-generator, such as the Borel-construction space $\mathrm{KU}_0 \!\sslash\! B\mathrm{U}(1)$ that classifies 3-twisted topological K-theory (cf. \cite[\S 4.1]{SS25-Flux}):
$$
  \begin{tikzcd}
    \mathrm{KU}_0
    \ar[r]
    &
    \mathrm{KU}_0
    \!\sslash\!
    B\mathrm{U}(1)
    \ar[
      d,
      ->>
    ]
    \\
    &
    B^2 \mathrm{U}(1)
  \end{tikzcd}
  \qquad\xrsquigarrow{\footnotesize \color{darkblue} Chern character\;}
  % \;\;\;\;\;
  % \overset{
  %   \scalebox{.7}{
  %     Chern character
  %   }
  % }{
  % \rightsquigarrow
  % }
  % \;\;\;\;\;
  \qquad 
  \def\arraystretch{1.2}
  \begin{array}{ccl}
    \mathrm{d}\, F_{2\bullet}
    &=&
    H_3\, F_{2\bullet-2}
    \\
    \\
    \mathrm{d}\, H_3 &=& 0
    \mathrlap{\,.}
  \end{array}
$$
The construction of such {\it characters} generalizes \cite{FSS23-Char} to non-abelian cohomology theories:

\medskip

\noindent
{\bf M-Brane charge in Cohomotopy.} Among the admissible flux-quantization laws for the C-field sourced by M-branes turns out to be
the most fundamental and most ancient non-abelian cohomology theory there is, known as (unstable) ``co-homotopy'' (since its classifying spaces are nothing but spheres), introduced by Pontrjagin in the 1930s (and later baptized by Spanier).

Concretely, the real-homotopy groups of $S^4$ have a generator in degree 4 (the identity map) and in degree 7 (the quaternionic Hopf fibration $S^7 \xrightarrow{h_{\mathbb{H}}} S^4$), while the real-cohomology only has a generator $G_4$ in degree 4. Hence $G_4$ must be closed while the other generator $G_7$ must be a coboundary for the otherwise induced cohomology class of $G_4\, G_4$. This way, the character map on 4-Cohomotopy reproduces the equation of motion \eqref{ElectricGaussLaw}
of the 11D SuGra C-field \cite{Sati13}\cite{SS24-PhaseSpace}:
\begin{equation}
  \label{CharacterIn4Cohomotopy}
  S^4
  % \;\;\;\;\;
  % \overset{
  %   \scalebox{.7}{character}
  % }{
  % \rightsquigarrow
  % }
  % \;\;\;\;\;
  \qquad 
  \xrsquigarrow{\footnotesize \color{darkblue} character}
  \qquad 
  \def\arraystretch{1.3}
  \begin{array}{ccl}
    \mathrm{d}\, G_7
    &=&
    \tfrac{1}{2}
    G_4\, G_4
    \\
    \mathrm{d}\,
    G_4 &=& 0 \;.
  \end{array}
\end{equation}

Careful analysis shows that assuming (``Hypothesis H'') 11D supergravity to be globally completed by demanding the C-field flux densities to be quantized in (tangentially twisted) 4-Cohomotopy
provably implies various subtle topological effects that are expected in M-theory \cite{FSS20-H}\cite{FSSHopf}\cite{FSS20TwistedString}\cite{SS21-M5Anomaly}, notably the condition that the sum of $G_4$ with $1/4$th of the first Pontrjagin form of the spin-connection is integral \cite[Prop. 3.13]{FSS20-H}.

\medskip

\noindent
{\bf 3-Form flux on M5-Branes.} 
Moreover, given an M5-brane $\Sigma^{1,5} \xhookrightarrow{\,\phi\,} X^{1,10}$ probing the bulk spacetime $X^{1,10}$, its worldvolume $\Sigma^{1,5}$ famously (but quite  \cite{HSW97}\cite{GSS24-FluxOnM5}) carries itself a non-linearly self-dual 3-flux density $H_3$ (sourced by string-like solitons inside the M5), satisfying the Bianchi identity
% \footnote{
% The same kind of Bianchi identity \eqref{H3Bianchi}
% appears on the M-theoretic correspondence space $\widehat{M}^{528} \xrightarrow{\;\phi\;} X^{11}$ (cite).
% }
\begin{equation}
  \label{H3Bianchi}
  \mathrm{d}
  \,
  H_3
  \;=\;
  \phi^\ast G_4
  \,,
\end{equation}
which, while nominally linear, inherits the non-linearity
\eqref{ElectricGaussLaw} of the source term $G_4$ on the right. An admissible non-abelian flux quantization law for the combination of \eqref{H3Bianchi} with \eqref{ElectricGaussLaw} turns out to be (tangentially twisted) 7-Cohomotopy {\it relative to} the bulk 4-Cohomotopy (where \eqref{H3Bianchi} reflects the vanishing of the class of the volume form of $S^4$ upon pullback to $S^7$).
This means that where the latter has as classifying space the 4-sphere, the former has as classifying space the 3-sphere {\it fibers} of the quaternionic Hopf fibration $\phi_{\mathbb{H}}$ \cite[\S 3.7]{FSS20-H}:
\begin{equation}
  \label{QuaternionicHopfFibration}
  \adjustbox{raise=5pt}{
  \begin{tikzcd}[
    row sep=18pt
  ]
    S^3
    \ar[
      r
    ]
    &
    S^7
    \ar[
      d,
      ->>,
      "{
        \phi_{\mathbb{H}}
      }"{pos=.4}
    ]
    &[-20pt]
    \simeq
    &[-20pt]
    S(\mathbb{H}^2)
    \ar[
      d,
      ->>,
      "{
        \mathrm{mod}
        \,
        \mathbb{H}^\times
      }"{pos=.3}
    ]
    \\
    & S^4
    &\simeq&
    \mathbb{H}P^1
  \end{tikzcd}
  }
  % \;\;\;\;\;
  % \overset{
  %   \scalebox{.7}{character}
  % }{
  % \rightsquigarrow
  % }
  % \;\;\;\;\;
  \qquad 
  \xrsquigarrow{\footnotesize \color{darkblue} character}
  \qquad
  \begin{array}{ccl}
    \mathrm{d}\, H_3
    &=&
    \phi^\ast G_4
    \\
    \\
    \mathrm{d}\, G_7 
    &=&
    \tfrac{1}{2} G_4 \, G_4
    \\
    \mathrm{d}\, G_4
    &=&
    0
    \mathrlap{\,.}
  \end{array}
\end{equation}

\medskip

\noindent
{\bf Gauge field on $A_1$-singularities.} More generally, for an M5-brane probing its would-be black brane horizon, namely probing
an $A_1$-type orbi-singularity of spacetime (i.e., locally the fixed locus of the $\ZTwo \subset \mathrm{Sp}(1)$-action on a patch $X^7 \times \mathbb{H} \subset X^{11}$) a further flux density $F_2$ appears (e.g. \cite[p. 92]{Shimizu18}, cf. \cite{SS20-EquivariantTwistorial}) and modifies \eqref{H3Bianchi} to
\begin{equation}
  \label{BianchiWithF2}
  \mathrm{d}
  \,
  H_3
  \;=\;
  \phi^\ast G_4
  \,+\,
  F_2 \, F_2
  \,.
\end{equation}
For vanishing $\phi^\ast G_4$ this relation is of the same form as \eqref{ElectricGaussLaw} and hence readily seen to be flux-quantized by the 2-sphere. Further inspection \cite{FSS22-GS} shows that in general \eqref{BianchiWithF2} is flux-quantized by the 2-sphere fibration over the 4-sphere that is also known as the {\it twistor fibration}, whose total space is $\mathbb{C}P^3$: 
\begin{equation}
  \label{TwistorFibration}
  \adjustbox{raise=5pt}{
  \begin{tikzcd}[
    row sep=18pt
  ]
    S^2
    \ar[
      r
    ]
    &
    \mathbb{C}P^3
    \ar[
      d,
      ->>,
      "{
        \phi_{\mathbb{C}}
      }"{pos=.4}
    ]
    &[-20pt]
    \simeq
    &[-20pt]
    S(\mathbb{H}^2)/
    S(\mathbb{C})
    \ar[
      d,
      ->>,
      "{
        \mathrm{mod}
        \,
        \mathbb{H}^\times
      }"{pos=.3}
    ]
    \\
    & S^4
    &\simeq&
    \mathbb{H}P^1
  \end{tikzcd}
  }
  % \;\;\;\;\;
  % \overset{
  %   \scalebox{.7}{character}
  % }{
  % \rightsquigarrow
  % }
  % \;\;\;\;\;
    \qquad 
  \xrsquigarrow{\footnotesize \color{darkblue} character}
  \qquad
  \begin{array}{ccl}
    \mathrm{d}\, H_3
    &=&
    \phi^\ast G_4
    \,+\,
    F_2\, F_2
    \\
    \\
    \mathrm{d}\, G_7 
    &=&
    \tfrac{1}{2} G_4 \, G_4
    \\
    \mathrm{d}\, G_4
    &=&
    0
    \mathrlap{\,.}
  \end{array}
\end{equation}

\medskip

\medskip

\noindent
{\bf Anyonic solitons in 2-Cohomotopy.} Now something remarkable happens: A deep theorem by Segal (\cite{Segal73}, cf. \cite[\S 4.1]{Kallel24}) shows that the moduli space of codimension=2 solitons\footnote{
The subscript $(-)_{\mathrm{cpt}}$ on a worldvolume domain --- as in \eqref{SegalTheorem} and \eqref{PuncturedM5Worldvolume} --- denotes its {\it one-point compactification}, reflecting the characteristic condition that solitonic charges {\it vanish at infinity}, cf. \cite[pp. 7, 14, 43]{SS23-Mf}\cite[\S 2.2]{SS25-Flux}.
} \cite[\S 2.2]{SS25-Flux}
 sourcing flux that is quantized in the 2-Cohomotopy \eqref{TwistorFibration}
have moduli space the pointed mapping space 
\begin{equation}
  \label{SegalTheorem}
  \mathrm{Maps}\big(
    \mathbb{R}^2_{\mathrm{cpt}}
    ,\,
    S^2
  \big)
  \;\;
  \simeq
  \;\;
  \mathbb{G}\mathrm{Conf}(\mathbb{R}^2)
  \,,
\end{equation}
equivalent to the ``group completion'' $\mathbb{G}$ of the {\it configuration space} $\mathrm{Conf}$ of points in the plane $\mathbb{R}^2$ (i.e. in the transverse space to the codim=2 solitons, in which they appear as points -- but (e.g. \cite{GM11}\cite{Williams20})
\begin{equation}
  \label{ConfClassifiesBraidGroup}
  \mathrm{Conf}(\mathbb{R}^2) 
  \;\simeq\; 
  \underset{n \in \mathbb{N}}{\sqcup}\; B \mathrm{Br}(n)
\end{equation}
is the classifying space for the {\it braid groups} $\mathrm{Br}(n)$ of motion of $n$ anyons in the plane!

On this, the 
``group completion'' $\mathbb{G}$ says essentially (\cite[p. 6]{SS24-AbAnyons}) that, besides the solitons that appear as points, there may also be {\it anti-solitons} that appear as points carrying a negative unit charge. 
This means that loops $
\ell \,\in\, \Omega \, \mathbb{G}\mathrm{Conf}(\mathbb{R}^2)$  describe just the kind of processes traditionally envisioned in discussion of topological quantum computation, where anyon/anti-anyon pairs are created out of the vacuum, then moved around each other, to eventually pair-annihilate again into the vacuum -- whereby their worldlines form knots and generally links.

In fact, careful analysis \cite[\S 6]{SS24-AbAnyons} shows that these loop processes in \eqref{SegalTheorem} are {\it framed} links (e.g. \cite[p. 15]{Ohtsuki01})
and that homotopy classes of these processes are the cobordism classes $[L]$ of these framed links, and that these are classified by
their total linking number $\# L$, including the framing number (cf. \hyperlink{FigureFramedLinks}{Fig. 1}):
\begin{equation}
  \label{FramedLinks}
  \begin{tikzcd}
  \pi_1
  \big(
    \mathbb{G}
    \,
    \mathrm{Conf}(\mathbb{R}^2)
  \big)
  \;\;
  \simeq
  \;\;
  \big\{\!\!
    \scalebox{.9}{
    Framed links
    }
  \!\!\big\}_{/\mathrm{cobordism}}
  \ar[
    rr,
    "{
      \#
    }",
    "{ \sim }"{swap}
  ]
  &&
  \mathbb{Z}
  \,.
  \end{tikzcd}
\end{equation}

\smallskip 
\begin{tabular}{cc}
\hypertarget{FigureFramedLinks}{}
\begin{tabular}{p{3.9cm}}
  \footnotesize
  {\bf Figure 1.} Some framed links $L$ with the framed unknots that they are cobordant to. The number $\# L$ \eqref{FramedLinks}
  is the sum of the linking- and self-linking (framing) number.\end{tabular}
&
\def\arraystretch{1.3}
\begin{tabular}{|cccc|}
\hline
{\bf Framed link}
&
cobordism
&
{\bf Framed unknot}
&
$\#L$
\\
\hline
\hline
&&&
\\[-13pt]
\adjustbox{
  raise=-1cm,
  scale=.5
}{
\begin{tikzpicture}[
  scale=1
]

\begin{scope}[
  shift={(1,0)}
]
\draw[line width=2, -Latex]
  (0:1) arc (0:180:1);
\end{scope}

\draw[line width=7,white]
  (0:1) arc (0:180:1);
\draw[line width=2, -Latex]
  (0:1) arc (0:180:1);

\draw[line width=2, -Latex]
  (180:1) arc (180:360:1);

\begin{scope}[shift={(1,0)}]
\draw[line width=7, white]
  (180:1) arc (180:360:1);
\draw[line width=2, -Latex]
  (180:1) arc (180:360:1);
\end{scope}

\node[gray]
  at (.5,.64) {\color{red} 
    \scalebox{.9}{$-$}
  };
\node[gray]
  at (.5,-.64) {\color{red}
    \scalebox{.9}{$-$}
  };
\end{tikzpicture}}
&
 \begin{tikzpicture}[decoration=snake]
   \draw[decorate, ->]
    (-.01,0) -- (0.54,0);
   \draw[decorate, ->]
    (0.01,0) -- (-0.54,0);
 \end{tikzpicture}
&
\adjustbox{
  raise=-1cm,
  scale=.6
}{
\begin{tikzpicture}

\begin{scope}[shift={(1,0)}]
\draw[line width=2, -Latex]
  (0:1) arc (0:180:1);
\end{scope}

\draw[line width=7,white]
  (0:1) arc (0:180:1);
\draw[line width=2, -Latex]
  (0:1) arc (0:180:1);

\draw[line width=2, -Latex]
  (180:1) arc (180:360:1);

\begin{scope}[shift={(1,0)}]
\draw[line width=7, white]
  (180:1) arc (180:360:1);
\draw[line width=2, -Latex]
  (180:1) arc (180:360:1);
\end{scope}

\draw[white,fill=white]
 (-.3,-.55) rectangle 
 (1.2,+.55);

\draw[line width=2]
  (.067,.55) 
  .. controls
    (.05-.2, .1) and 
    (.86+.2, .1) ..
  (+.84,.55);

\begin{scope}[
  shift={(.1,0)},
  yscale=-1
]
\draw[line width=2]
  (.067,.55) 
  .. controls
    (.05-.2, .1) and 
    (.86+.2, .1) ..
  (+.84,.55);
\end{scope}
 
\end{tikzpicture}}
&
$-2$
\\
&&&
\\[-12pt]
\adjustbox{
  raise=-2.6cm,
  scale=.5
}{
\begin{tikzpicture}
\foreach \n in {0,1,2} {
\begin{scope}[
  rotate=\n*120-4
]
\draw[
  line width=2,
  -Latex
]
 (0,-1)
   .. controls
   (-1,.2) and (-2,2) ..
 (0,2)
   .. controls
   (1,2) and (1,1) ..
  (.9,.7);
\end{scope}

\node[darkgreen]
  at (\n*120+31:.7) {
    \scalebox{1}{$+$}
  };
};
\end{tikzpicture}
}
&
 \begin{tikzpicture}[decoration=snake]
   \draw[decorate, ->]
    (-.01,0) -- (0.54,0);
   \draw[decorate, ->]
    (0.01,0) -- (-0.54,0);
 \end{tikzpicture}
&
\adjustbox{
  raise=-2.6cm,
  scale=.5
}{
\begin{tikzpicture}
\foreach \n in {0,1,2} {
\begin{scope}[
  rotate=\n*120-4
]
\draw[
  line width=2,
  -Latex
]
 (0,-1)
   .. controls
   (-1,.2) and (-2,2) ..
 (0,2)
   .. controls
   (1,2) and (1,1) ..
  (.9,.7);
\end{scope}
};

\draw[white,fill=white]
  (-.9,-.7)
  rectangle
  (.9,.2);

\draw[line width=2]
  (-.79,.21) 
    .. controls
    (-.7,-.25) and (+.7,-.25) ..
  (.79,.21);

\draw[line width=2]
  (-.27,-.7) 
  .. controls
    (-.4,-.2) and 
    (+.4,-.2) ..
  (+.27,-.7);

\draw[white,fill=white]
 (-.5,.8) 
 rectangle (+.5,-.2);

\draw[line width=2]
  (-.5,.58)
  .. controls
    (-.0,.7) and (-.0,-.26)
    ..
  (-.5,-.06);

\draw[line width=2]
  (+.5,.58)
  .. controls
    (+.0,.7) and (+.0,-.26)
    ..
  (+.5,-.06);

\end{tikzpicture}
}
&
3
\\[-11pt]
\adjustbox{
  raise=-1cm,
  scale=.8
}{
\begin{tikzpicture}

\node at (0,0) {
  \includegraphics[width=2.2cm]{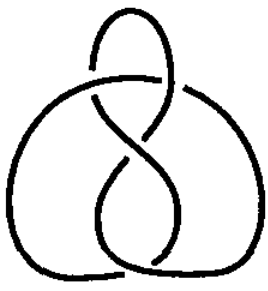}
};

\draw[line width=1.2pt,-Latex]
  (-.276,-.8) -- 
  (-.274,-.8-.02);
\draw[line width=1.2pt,-Latex]
  (.35, -.5) --
  (.35, -.47);
\draw[line width=1.2pt,-Latex]
  (-.09,1.09) --
  (-.09+.05, 1.13);
\draw[line width=1.2pt,-Latex]
  (-.18,.54) --
  (-.19-.01, .54);

\node[red]
  at (0,-1.2) {
    \scalebox{.7}{$-$}
  };
\node[red]
  at (.3,0) {
    \scalebox{.7}{$-$}
  };
\node[darkgreen]
  at (.5,.64) {
    \scalebox{.7}{$+$}
  };
\node[darkgreen]
  at (-.57,.64) {
    \scalebox{.7}{$+$}
  };

\end{tikzpicture}
}
&
 \begin{tikzpicture}[decoration=snake]
   \draw[decorate, ->]
    (-.01,0) -- (0.54,0);
   \draw[decorate, ->]
    (0.01,0) -- (-0.54,0);
 \end{tikzpicture}
&
\adjustbox{
  scale=.8,
  raise=-.8cm
}{
\begin{tikzpicture}

\node at (0,0) {
  \includegraphics[width=2.2cm]{FigureEightKnot.png}
};

\draw[white,fill=white]
  (-.4,-.8) rectangle (.5,-.3);

\draw[line width=1.5]
  (-.27,-.81)
    ..
    controls
    (-.32, -.6) and (.4,-.6) ..
  (.3,-.81);

\begin{scope}[
  shift={(0,-1.1)},
  yscale=-1
]
\draw[line width=1.5]
  (-.19,-.81)
    ..
    controls
    (-.3, -.6) and (.4,-.6) ..
  (.28,-.81);
\end{scope}

\draw[white,fill=white]
  (-.2,.4) rectangle (+.15,1.4);

\draw[line width=1.5]
 (-.22,1.01) 
 .. controls 
   (0,1.1) and (0,.55) ..
 (-.22,.54);

\begin{scope}[
  shift={(-.03,0)},
  xscale=-1
]
\draw[line width=1.5]
 (-.22,1) 
 .. controls 
   (0,1.1) and (0,.55) ..
 (-.22,.54);
\end{scope}

\draw[
  line width=1.2,
  -Latex
]
  (-1.01,-.6) --
  (-1.01,-.6-.01);

\end{tikzpicture}
}
&
0
\\
\hline
\end{tabular}
\\
\end{tabular}

\vspace{.4cm}

\noindent
{\bf Quantum observables on quantized fluxes.}
Once supergravity is completed by a flux-quantiz-ation law $\mathcal{A}$ this way,
then \cite{SS23-Obs}
for every spacetime- or worldvolume-domain, its mapping space into $\mathcal{A}$ constitutes the moduli space of topological sectors of solitonic higher gauge field configurations, being the higher analog of configurations of Abrikosov vortices in electromagnetism. When the domain is a principal bundle $\Sigma^{1,5} \xrightarrow{\;} \Sigma^{1,4}$ of circle fibers (such as the  ``M-theory circle'') then the quantum observables on the topological sectors of such flux-quantized fields form the Ponrjagin homology algebra (cf. \cite[\S 3]{SS23-Obs}) of the loop space of this moduli space:
\vspace{1mm} 
\begin{equation}
  \label{QuantumObservables}
  \scalebox{.7}{
    \color{darkblue}
    \bf
    \def\arraystretch{.9}
    \begin{tabular}{c}
      Topological
      quantum observables
      \\
      of $\mathcal{A}$-quantized
      fluxes
      \\
      on M5-worldvolume $\Sigma^{1,4} \times S^1$
    \end{tabular}
  }
  \mathrm{Obs}_\bullet
  \;
  =
  \;
  H_\bullet\big(
    \Omega
    \,
    \mathrm{Maps}(
      \Sigma^{1,4}
      ,\,
      \mathcal{A}
    )
    ;\,
    \mathbb{C}
  \big)
  \,.
\end{equation}
This is the non-perturbative quantization of a small but crucial fragment of super-gravity, rather complementary to the traditional focus of interest: Instead of local quantum effects such as of graviton scattering visible on any coordinate chart, here we deal with the global topological quantum effects. 

\medskip

\noindent
{\bf Anyonic quantum observables on M5-branes.} Concretely, consider a worldvolume domain
on which to measure the charge of 3-brane solitons inside the M5-brane (cf. \cite[\S 14.6.1]{West12}\cite[p. 26]{SS23-DefectBranes}), wrapped over the M-theory circle,
\begin{equation}
  \label{M5Worldvolume}
  \grayoverbrace{
    \Sigma^{1,5}
  }{
    \mathclap{
    \scalebox{.7}{
    5-brane worldvolume
    }
    }
  }
  \;\; 
    =
  \;\;
  \grayunderbrace{
  \mathbb{R}^{1,2}
  \times
  S^1
  }{
    \mathclap{
    \scalebox{.7}{
      3-brane worldvolume
    }
    }
  }
  \;
  \times
  \;
  \grayoverbrace{
  \mathbb{R}^{2}_{\mathrm{cpt}}
  }{
    \mathclap{
      \scalebox{.7}{
        transverse space
      }
    }
  }
\end{equation}
in, for simplicity, a background with vanishing C-field.
Then, for any choice of admissible flux quantization law $\mathcal{A}$, the topological quantum observables on these 3-brane solitons is given by \eqref{QuantumObservables}, which for the choice \eqref{TwistorFibration} is
\cite[\S 4]{SS24-AbAnyons}
the group algebra of cobordism classes of framed links, under their connected sum:
\begin{equation}
  \label{QuantumObservables}
  \def\arraystretch{1.7}
  \begin{array}{ccll}
    \mathrm{Obs}_0
    &\defneq&
    H_0\Big(
      \Omega\,
      \mathrm{Maps}\big(
        \mathbb{R}^{1,2}
        \times
        \mathbb{R}^2_{\mathrm{cpt}}
        ,\,
        S^2
      \big)
      ;\,
      \mathbb{C}
    \Big)
    &
    \proofstep{
      by
      \eqref{QuantumObservables}
    }
    \\
    &\defneq&
    H_0\Big(
      \Omega\,
      \mathrm{Maps}\big(
        \mathbb{R}^2_{\mathrm{cpt}}
        ,\,
        S^2
      \big)
      ;\,
      \mathbb{C}
    \Big)
    &
    \proofstep{
      since $\mathbb{R}^{1,2}$
      is cntrctbl
    }
    \\
    &\defneq&
    H_0\Big(
      \Omega\,
      \mathbb{G}
      \,
      \mathrm{Conf}(\mathbb{R}^2)
      ;\,
      \mathbb{C}
    \Big)   
    &
    \proofstep{
      by
      \eqref{SegalTheorem}
    }
    \\
    &\defneq&
    \mathbb{C}\Big[
      \pi_1
      \big(
      \mathbb{G}
      \mathrm{Conf}(\mathbb{R}^2)
      \big)
    \Big]
    &
    \proofstep{
      by 0-Hurewicz
    }
    \\
    &\defneq&
    \mathbb{C}\Big[
      \big\{\!
        \scalebox{.8}{
          Framed links
        }
      \!\big\}_{/\mathrm{cbrdsm}}
    \Big]
    &
    \proofstep{
      by
      \eqref{FramedLinks}
      .
    }
  \end{array}
\end{equation}

\medskip

\noindent
{\bf Anyonic quantum states on M5-branes.} This implies, by the rules of algebraic quantum theory, that \cite[Prop. 4.3]{SS24-AbAnyons} the corresponding {\it pure} quantum states $\vert \psi \rangle$ are, via the expectation values that they induce on the observables \eqref{QuantumObservables}, the algebra homomorphisms 
\begin{equation}
  \label{PhaseFactor}
  \begin{tikzcd}[row sep=-5pt, column sep=0pt]
    \mathbb{C}\Big[
      \big\{\!\!
        \scalebox{.9}{
          Framed links
        }
      \!\! \big\}_{/\mathrm{cbrdsm}}
    \Big]
    \ar[
      rr,
      "{
        \langle 
          k
        \vert
        -
        \vert
          k
        \rangle
      }"
    ]
    &&
    \mathbb{C}
    \\
    {[L]}
    &\longmapsto&
    \exp\big(
      \tfrac{
        2 \pi \mathrm{i}
      }{
        k
      }
      \# L
    \big)
    \,,
  \end{tikzcd}
\end{equation}
which are generated by states $\vert k \rangle$ for $k \in \mathbb{Z}$, as shown.
These are exactly the traditional\footnote{
In traditional discussion of these observables, the framing on the links and the inclusion of the self-linking number are introduced in an {\it ad hoc} manner in order to work around an otherwise ill-defined term obtained by path-integral heuristics. In contrast, in our derivation above these features emerge by rigorous analysis of quantum observables of the flux-quantized self-dual higher gauge field.
} quantum observables of $\mathrm{U}(1)$ Chern-Simons theory as expected for abelian anyons! 
It is believed that such quantum states have been observed \cite{NLGM20} in fractional quantum Hall (FQH) systems \cite{Stormer99}\cite{Girvin04}. 

\medskip

However, as may often be  overlooked, anyonic states in this form are not yet useful for quantum computation: While the anyonic braiding statistics is visible in the phase factor \eqref{PhaseFactor}, there is no control yet over the movement of these anyons around each other in order to implement topological quantum gate operations (cf. \cite[\S 3]{MySS24}).

We next see how this control arises in our holographic theory.

\medskip

\medskip

\noindent
{\bf Topological quantum gates.}
Namely, to model classically controllable anyonic {\it defects} in addition to the above anyon ``virtual particles'', consider deleting a subset $\mathbf{n} \subset \mathbb{R}^2$ of $n \in \mathbb{N}$ defect points from the transverse plane (just as one deletes the singular locus of a black hole or black brane defect from the spacetime domain, cf. \cite[\S 2.2]{SS25-Flux}) and take the transverse space of the 3-brane soliton inside the M5-brane now to be the $\mathbf{n}$-punctured plane, generalizing \eqref{M5Worldvolume} to
\begin{equation}
  \label{PuncturedM5Worldvolume}
  \grayoverbrace{
    \Sigma^{1,5}
  }{
    \mathclap{
    \scalebox{.7}{
    5-brane worldvolume
    }
    }
  }
  \;\; 
    =
  \;\;
  \grayunderbrace{
  \mathbb{R}^{1,2}
  \times
  S^1
  }{
    \mathclap{
    \scalebox{.7}{
      3-brane worldvolume
    }
    }
  }
  \;
  \times
  \;
  \grayoverbrace{
  \big(
    \mathbb{R}^{2}
    \setminus
    \mathbf{n}
  \big)_{\mathrm{cpt}}
  }{
    \mathclap{
      \scalebox{.7}{
        transverse space
      }
    }
  }
  \,.
\end{equation}
The topological symmetries of the worldvolume domain fixing the 3-brane locus 
is the mapping class group of $\mathbb{R}^2 \setminus \mathbf{n}$ fixing the original point at infinity, which in turn is the braid group $\mathrm{Br}_n$ (quotiented by its center, cf. \cite[\S 1.4]{GM11}) which as such canonically acts on the resulting quantum observables, formed as above in \eqref{QuantumObservables}
$$
  \mathrm{Obs}_\bullet
  \;\;
  :=
  \;\;
  H_\bullet
  \Big(
  \Omega
  \,
  \mathrm{Maps}\big(
    (
      \mathbb{R}^2
      \setminus
      \mathbf{n}
    )_{\mathrm{cpt}}
    ,\,
    S^2
  \big)
  ;\,
  \mathbb{C}
  \Big)
  \,.
$$
But this in turn gives an action of $\mathrm{Br}_n$ on the corresponding Hilbert space of quantum states. This is what counts as a set of topological quantum gates, where an adiabatic braid-motion of anyon defects around each other acts by quantum phases on the system's Hilbert space.

\begin{tikzpicture}

  \shade[right color=lightgray, left color=white]
    (3,-3)
      --
      node[above, yshift=-1pt, 
      xshift=14pt,
      sloped]{
        \scalebox{.7}{
          \color{darkblue}
          \bf
          transverse space
        }
      }
    (-1,-1)
      --
    (-1.21,1)
      --
    (2.3,3);

  \draw[]
    (3,-3)
      --
    (-1,-1)
      --
    (-1.21,1)
      --
    (2.3,3)
      --
    (3,-3);

\draw[-Latex]
  ({-1 + (3+1)*.3},{-1+(-3+1)*.3})
    to
  ({-1 + (3+1)*.29},{-1+(-3+1)*.29});

\draw[-Latex]
    ({-1.21 + (2.3+1.21)*.3},{1+(3-1)*.3})
      --
    ({-1.21 + (2.3+1.21)*.29},{1+(3-1)*.29});

\draw[-Latex]
    ({2.3 + (3-2.3)*.5},{3+(-3-3)*.5})
      --
    ({2.3 + (3-2.3)*.49},{3+(-3-3)*.49});

\draw[-latex]
    ({-1 + (-1.21+1)*.53},{-1 + (1+1)*.53})
      --
    ({-1 + (-1.21+1)*.54},{-1 + (1+1)*.54});

  \begin{scope}[rotate=(+8)]
   \draw[dashed]
     (1.5,-1)
     ellipse
     ({.2*1.85} and {.37*1.85});
   \begin{scope}[
     shift={(1.5-.2,{-1+.37*1.85-.1})}
   ]
     \draw[->, -Latex]
       (0,0)
       to
       (180+37:0.01);
   \end{scope}
   \begin{scope}[
     shift={(1.5+.2,{-1-.37*1.85+.1})}
   ]
     \draw[->, -Latex]
       (0,0)
       to
       (+37:0.01);
   \end{scope}
   \begin{scope}[shift={(1.5,-1)}]
     \draw (.43,.65) node
     { \scalebox{.8}{$
     $} };
  \end{scope}
  \draw[fill=white, draw=gray]
    (1.5,-1)
    ellipse
    ({.2*.3} and {.37*.3});
  \draw[line width=3.5, white]
   (1.5,-1)
   to
   (-2.2,-1);
  \draw[line width=1.1]
   (1.5,-1)
   to node[
     above, 
     yshift=-4pt, 
     pos=.85]{
     \;\;\;\;\;\;\;\;\;\;\;\;\;
     \rotatebox[origin=c]{7}
     {
     \scalebox{.7}{
     \color{darkorange}
     \bf
     \colorbox{white}{defect anyon}
     }
     }
   }
   (-2.2,-1);
  \draw[
    line width=1.1
  ]
   (1.5+1.2,-1)
   to
   (3.5,-1);
  \draw[
    line width=1.1,
    densely dashed
  ]
   (3.5,-1)
   to
   (4,-1);

  \draw[line width=3, white]
   (-2,-1.3)
   to
   (0,-1.3);
  \draw[-latex]
   (-2,-1.3)
   to
   node[
     below, 
     yshift=+3pt,
     xshift=-7pt
    ]{
     \scalebox{.7}{
       \rotatebox{+7}{
       \color{darkblue}
       \bf
       parameter
       }
     }
   }
   (0,-1.3);
  \draw[dashed]
   (-2.7,-1.3)
   to
   (-2,-1.3);

 \draw
   (-3.15,-.8)
   node{
     \scalebox{.7}{
       \rotatebox{+7}{
       \color{darkgreen}
       \bf
       braiding
       }
     }
   };

  \end{scope}

  \begin{scope}[shift={(-.2,1.4)}, scale=(.96)]
  \begin{scope}[rotate=(+8)]
  \draw[dashed]
    (1.5,-1)
    ellipse
    (.2 and .37);
  \draw[fill=white, draw=gray]
    (1.5,-1)
    ellipse
    ({.2*.3} and {.37*.3});
  \draw[line width=3.1, white]
   (1.5,-1)
   to
   (-2.3,-1);
  \draw[line width=1.1]
   (1.5,-1)
   to
   (-2.3,-1);
  \draw[line width=1.1]
   (1.5+1.35,-1)
   to
   (3.6,-1);
  \draw[
    line width=1.1,
    densely dashed
  ]
   (3.6,-1)
   to
   (4.1,-1);
  \end{scope}
  \end{scope}

  \begin{scope}[shift={(-1,.5)}, scale=(.7)]
  \begin{scope}[rotate=(+8)]
  \draw[dashed]
    (1.5,-1)
    ellipse
    (.2 and .32);
  \draw[fill=white, draw=gray]
    (1.5,-1)
    ellipse
    ({.2*.3} and {.32*.3});
  \draw[line width=3.1, white]
   (1.5,-1)
   to
   (-1.8,-1);
\draw
   (1.5,-1)
   to
   (-1.8,-1);
  \draw
    (5.23,-1)
    to
    (6.4-.6,-1);
  \draw[densely dashed]
    (6.4-.6,-1)
    to
    (6.4,-1);
  \end{scope}
  \end{scope}

\draw (1.73,-1.06) node
 {
  \scalebox{.8}{
    $k_{{}_{I}}$
  }
 };

\begin{scope}
[ shift={(-2,-.55)}, rotate=-82.2  ]

 \begin{scope}[shift={(0,-.15)}]

  \draw[]
    (-.2,.4)
    to
    (-.2,-2);

  \draw[
    white,
    line width=1.1+1.9
  ]
    (-.73,0)
    .. controls (-.73,-.5) and (+.73-.4,-.5) ..
    (+.73-.4,-1);
  \draw[
    line width=1.1
  ]
    (-.73+.01,0)
    .. controls (-.73+.01,-.5) and (+.73-.4,-.5) ..
    (+.73-.4,-1);

  \draw[
    white,
    line width=1.1+1.9
  ]
    (+.73-.1,0)
    .. controls (+.73,-.5) and (-.73+.4,-.5) ..
    (-.73+.4,-1);
  \draw[
    line width=1.1
  ]
    (+.73,0+.03)
    .. controls (+.73,-.5) and (-.73+.4,-.5) ..
    (-.73+.4,-1);

  \draw[
    line width=1.1+1.9,
    white
  ]
    (-.73+.4,-1)
    .. controls (-.73+.4,-1.5) and (+.73,-1.5) ..
    (+.73,-2);
  \draw[
    line width=1.1
  ]
    (-.73+.4,-1)
    .. controls (-.73+.4,-1.5) and (+.73,-1.5) ..
    (+.73,-2);

  \draw[
    white,
    line width=1.1+1.9
  ]
    (+.73-.4,-1)
    .. controls (+.73-.4,-1.5) and (-.73,-1.5) ..
    (-.73,-2);
  \draw[
    line width=1.1
  ]
    (+.73-.4,-1)
    .. controls (+.73-.4,-1.5) and (-.73,-1.5) ..
    (-.73,-2);

 \draw
   (-.2,-3.3)
   to
   (-.2,-2);
 \draw[
   line width=1.1,
   densely dashed
 ]
   (-.73,-2)
   to
   (-.73,-2.5);
 \draw[
   line width=1.1,
   densely dashed
 ]
   (+.73,-2)
   to
   (+.73,-2.5);

  \end{scope}
\end{scope}

\begin{scope}[shift={(-5.6,-.75)}]

  \draw[line width=3pt, white]
    (3,-3)
      --
    (-1,-1)
      --
    (-1.21,1)
      --
    (2.3,3)
      --
    (3, -3);

  \shade[right color=lightgray, left color=white, fill opacity=.7]
    (3,-3)
      --
    (-1,-1)
      --
    (-1.21,1)
      --
    (2.3,3);

  \draw[]
    (3,-3)
      --
    (-1,-1)
      --
    (-1.21,1)
      --
    (2.3,3)
      --
    (3, -3);

\draw (1.73,-1.06) node
 {
  \scalebox{.8}{
    $k_{{}_{I}}$
  }
 };

\draw[-Latex]
  ({-1 + (3+1)*.3},{-1+(-3+1)*.3})
    to
  ({-1 + (3+1)*.29},{-1+(-3+1)*.29});

\draw[-Latex]
    ({-1.21 + (2.3+1.21)*.3},{1+(3-1)*.3})
      --
    ({-1.21 + (2.3+1.21)*.29},{1+(3-1)*.29});

\draw[-Latex]
    ({2.3 + (3-2.3)*.5},{3+(-3-3)*.5})
      --
    ({2.3 + (3-2.3)*.49},{3+(-3-3)*.49});

\draw[-latex]
    ({-1 + (-1.21+1)*.53},{-1 + (1+1)*.53})
      --
    ({-1 + (-1.21+1)*.54},{-1 + (1+1)*.54});

  \begin{scope}[rotate=(+8)]
   \draw[dashed]
     (1.5,-1)
     ellipse
     ({.2*1.85} and {.37*1.85});
   \begin{scope}[
     shift={(1.5-.2,{-1+.37*1.85-.1})}
   ]
     \draw[->, -Latex]
       (0,0)
       to
       (180+37:0.01);
   \end{scope}
   \begin{scope}[
     shift={(1.5+.2,{-1-.37*1.85+.1})}
   ]
     \draw[->, -Latex]
       (0,0)
       to
       (+37:0.01);
   \end{scope}
  \draw[fill=white, draw=gray]
    (1.5,-1)
    ellipse
    ({.2*.3} and {.37*.3});
 \end{scope}

   \begin{scope}[shift={(-.2,1.4)}, scale=(.96)]
  \begin{scope}[rotate=(+8)]
  \draw[dashed]
    (1.5,-1)
    ellipse
    (.2 and .37);
  \draw[fill=white, draw=gray]
    (1.5,-1)
    ellipse
    ({.2*.3} and {.37*.3});
\end{scope}
\end{scope}

  \begin{scope}[shift={(-1,.5)}, scale=(.7)]
  \begin{scope}[rotate=(+8)]
  \draw[dashed]
    (1.5,-1)
    ellipse
    (.2 and .32);
  \draw[fill=white, draw=gray]
    (1.5,-1)
    ellipse
    ({.2*.3} and {.37*.3});
\end{scope}
\end{scope}

\begin{scope}
[ shift={(-2,-.55)}, rotate=-82.2  ]

 \begin{scope}[shift={(0,-.15)}]

 \draw[line width=3, white]
   (-.2,-.2)
   to
   (-.2,2.35);
 \draw
   (-.2,.5)
   to
   (-.2,2.35);
 \draw[dashed]
   (-.2,-.2)
   to
   (-.2,.5);

\end{scope}
\end{scope}

\begin{scope}
[ shift={(-2,-.55)}, rotate=-82.2  ]

 \begin{scope}[shift={(0,-.15)}]

 \draw[
   line width=3, white
 ]
   (-.73,-.5)
   to
   (-.73,3.65);
 \draw[
   line width=1.1
 ]
   (-.73,.2)
   to
   (-.73,3.65);
 \draw[
   line width=1.1,
   densely dashed
 ]
   (-.73,.2)
   to
   (-.73,-.5);
 \end{scope}
 \end{scope}

\begin{scope}
[ shift={(-2,-.55)}, rotate=-82.2  ]

 \begin{scope}[shift={(0,-.15)}]

 \draw[
   line width=3.2,
   white]
   (+.73,-.6)
   to
   (+.73,+3.7);
 \draw[
   line width=1.1,
   densely dashed]
   (+.73,-0)
   to
   (+.73,+-.6);
 \draw[
   line width=1.1 ]
   (+.73,-0)
   to
   (+.73,+3.71);
\end{scope}
\end{scope}

\end{scope}

\draw
  (-2.2,-4.2) node
  {
    \scalebox{1.2}{
      $
       \mathllap{
          \raisebox{1pt}{
            \scalebox{.58}{
              \color{darkblue}
              \bf
              \def\arraystretch{.9}
              \begin{tabular}{c}
                some quantum state for
                \\
                fixed defect positions
                \\
                $k_1, k_2, \cdots$
                at time
                {\color{purple}$t_1$}
              \end{tabular}
            }
          }
          \hspace{-5pt}
       }
        \big\vert
          \psi({\color{purple}t_1})
        \big\rangle
      $
    }
  };

\draw[|->]
  (-1.3,-4.1)
  to
  node[
    sloped,
    yshift=5pt
  ]{
    \scalebox{.7}{
      \color{darkgreen}
      \bf
      unitary adiabatic transport
    }
  }
  node[
    sloped,
    yshift=-5pt,
    pos=.4
  ]{
    \scalebox{.7}{
      }
  }
  (+2.4,-3.4);

\draw
  (+3.2,-3.85) node
  {
    \scalebox{1.2}{
      $
        \underset{
          \raisebox{-7pt}{
            \scalebox{.55}{
              \color{darkblue}
              \bf
              \def\arraystretch{.9}
               \begin{tabular}{c}
              another quantum state for
                \\
                fixed defect positions
                \\
                $k_1, k_2, \cdots$
                at time
                {\color{purple}$t_2$}
              \end{tabular}
            }
          }
        }{
        \big\vert
          \psi({\color{purple}t_2})
        \big\rangle
        }
      $
    }
  };

\end{tikzpicture}

An analogous analysis for a more sophisticated situation of {\it intersecting} M5-branes and resulting in {\it non-abelian} anyons was given in \cite{SS23-DefectBranes}.

\medskip

{\bf In summary} so far, this shows that the fundamentals of topological quantum logic gates, acting by adiabatic braiding of worldlines of anyonic defects, arise quite naturally from the non-perturbative quantization of the topological sector of solitons on single M5-branes in 11D supergravity, {\it if} flux-quantization is taken into account, of the bulk C-field and of the self-dual tensor field on the worldvolume, whose non-linear Gau{\ss} law \eqref{TwistorFibration}
is seen to reflect anyonic soliton charges in {\it non-abelian} generalized cohomology (concretely, in unstable Cohomotopy).

%%%%%%%%%%%%%%%%%%%%%%%%%%%%%%%%
\section{... and Computation}
\label{andComputation}
%%%%%%%%%%%%%%%%%%%%%%%%%%%%%%%

\noindent
{\bf Formulation in homotopically typed programming language.}
To bring out this relation between flux-quantized supergravity and (quantum) computation more manifestly, we may observe \cite{MySS24} that the elementary algebro-topological/homotopy-theoretic nature 
\cite{SS25-Flux}\cite{SS23-Obs} of quantum observables on flux-quantized fields --- as exhibited e.g. in \eqref{QuantumObservables} above  --- lends itself (exposition in \cite{Myers2024}) to formalized expression in novel {\it homotopically-typed} programming languages (cf.  \S\ref{Vista} and \cite{HoTTBook}, such as {\tt Agda} \cite{Escardo19} or {\tt cubicalAgda} \cite{MyersRiley24}) and better yet \cite{Sc14}\cite{SS23-Entanglement}\cite{SS23-Monadology}\cite{SS23-Reality} in languages with {\it linear homotopy types}, of which a prototype design has recently been described \cite{Riley22}.

To wit, the core mechanism of topological holonomic quantum gates \cite{ZanardiRasetti99}\cite{Zhang23}, parallel-transporting the quantum state of a system along paths of classical parameters (such as anyon defect positions) is, strikingly, {\it native} to such languages (``type transport'', cf. \cite[p. 39]{MySS24}\cite[\href{https://github.com/CQTS/introduction-to-cubical/blob/master/lectures/2--Paths-and-Identifications/2-5--Transport.lagda.md}{\S 2.5}]{MyersRiley24}):

\begin{center}
\includegraphics[width=13cm]{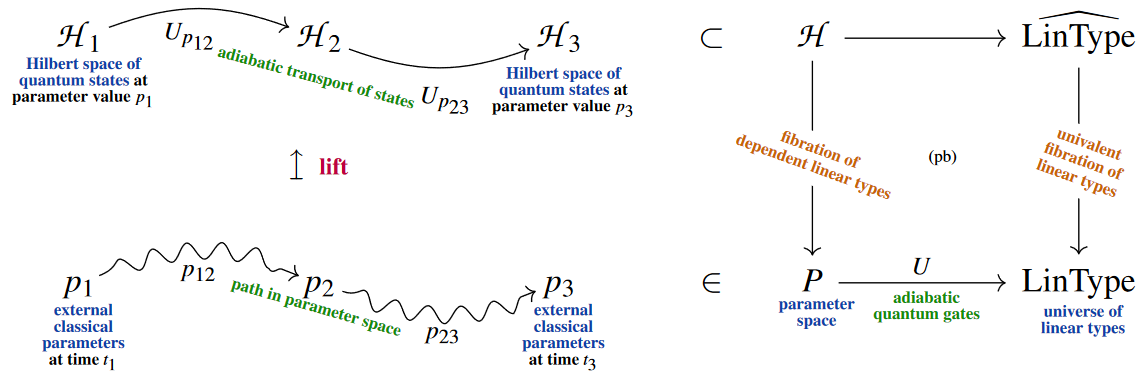}
\end{center}

Also native to homotopically-typed languages is the declaration of classifying spaces, such as for the braid group in \eqref{ConfClassifiesBraidGroup}, which means that their general knowledge of transport readily specializes to the case of braid gates:
\vspace{-2mm} 
\begin{center}
\begin{tikzpicture}[scale=.4]

\draw (-1,3.5) node {
  $\HilbertSpace{H}_1$
};
\draw
  [
    ->,
    bend left=20
  ] 
  (0,3.5) to
  node[
    yshift=6pt
  ]
  {
    \scalebox{.7}{
      \color{darkgreen}
      \bf 
      anyon braid quantum gate
    }
  }
  (10,3.5);
\draw (11,3.5) node {
  $\HilbertSpace{H}_2$
};

\draw
  (5,1.4)
  node
  {
    \scalebox{.7}{
      \color{darkgreen}
      \bf 
      path in config. space of defect points
    }
  };

\draw (17,3.4) node {
  \scalebox{1.2}{$\HilbertSpace{H}$}
};

\draw (17,-2) node {
  $\mathrm{Conf}(\mathbb{R}^2)$
};

\draw[->] (17,2.4) to (17,-.9);

\begin{scope}[shift={(0,0)}]
\rbraid
\end{scope}

\begin{scope}[shift={(0,-1.5)}]
 \strand
\end{scope}

\begin{scope}[shift={(0,-3)}]
  \lbraid
\end{scope}

\begin{scope}[shift={(0,-4.5)}]
 \strand
\end{scope}

\begin{scope}[shift={(2,.5)}]
  \strand
\end{scope}

\begin{scope}[shift={(2,-1)}]
  \lbraid
\end{scope}

\begin{scope}[shift={(2,-3)}]
  \lbraid
\end{scope}

\begin{scope}[shift={(2,-4.5)}]
 \strand
\end{scope}

\begin{scope}[shift={(4,0)}]
  \rbraid
\end{scope}

\begin{scope}[shift={(4,-2)}]
  \rbraid
\end{scope}

\begin{scope}[shift={(4,-4)}]
  \rbraid
\end{scope}

\begin{scope}[shift={(6,.5)}]
  \strand
\end{scope}

\begin{scope}[shift={(6,-1)}]
  \lbraid
\end{scope}

\begin{scope}[shift={(6,-3)}]
  \lbraid
\end{scope}

\begin{scope}[shift={(6,-4.5)}]
  \strand
\end{scope}

\begin{scope}[shift={(8,0)}]
  \rbraid
\end{scope}

\begin{scope}[shift={(8,-1.5)}]
  \strand
\end{scope}

\begin{scope}[shift={(8,-2.5)}]
  \strand
\end{scope}

\begin{scope}[shift={(8,-4)}]
  \rbraid
\end{scope}

\begin{scope}[shift={(10,0)}]
  \rbraid
\end{scope}

\begin{scope}[shift={(10,-2)}]
  \rbraid
\end{scope}

\begin{scope}[shift={(10,-4)}]
  \rbraid
\end{scope}

\end{tikzpicture}
\end{center}

Thereby, the encoding of topological quantum gates in a homotopically-typed programming language becomes essentially a 1-liner \cite[Thm. 6.8]{MySS24}.

This is remarkable: When topological quantum computers become a reality (or their quantum simulation becomes refined enough, cf. recent progress in \cite{ITV24}) their hardware-level quantum gates will be completely different from the idealized gates familiar from traditional qbit-based quantum circuits (Hadamard, CNOT, etc.) and efficient (hardware-aware) topological quantum programming languages will need to reflect this (cf. \cite[p. 3]{SS23-Monadology}). 

\medskip

{\bf In final conclusion} this means that the embedding (``geometric engineering'') of topological qbits into quantized topological sectors of  flux-quantized supergravity with M-brane probes illuminates both the quantum-physical as well as the quantum-information theoretic nature of anyons, both without relying on the unrealistic large-$N$ limit of existing holographic descriptions of quantum materials.

%%%%%%%%%%%%%%%%%%%%%%%
\section{Vista}
\label{Vista}
%%%%%%%%%%%%%%%%%%%%%%%%%%%

In reaction to and amplification of some of the thoughts of our editors expressed in \cite{AER23},
we close with more meta-physical remarks on the relevance of (cohesive) homotopy type theory in the foundations not just of \mbox{(quantum-)}computation but of fundamental physics and potentially of M-theory.

\medskip

\noindent 
{\bf Computation and physical process.}
In our age of electro-mechanical computers, and at the plausible dawn of a new age of quantum-mechanical computers, it is a truism that any {\it computation is a physical process}, possibly a very fundamental physical process (say, if we think of photonic quantum computation). The reverse of this truism, that possibly all {\it physical processes are computations}, hence that the history of the universe is the unfolding of an ancient primordial algorithm (\cite{Zuse69}\cite{Wolfram02}\cite{Zenil12}), is 
thought provoking --- all the more since it is less clear what it would actually mean.

\medskip

\noindent {\bf Computation and mathematical proof.}
On this issue we highlight that the field of mathematical logic has long developed an analogous relation: In the view of {\it constructive mathematics} (cf. \cite[\S 8]{Zenil12}, essentially what was originally called ``intuitionism'') the proof of a theorem must consist of the actual {\it construction} of a {\it witness} of its truth --- notably existential statements,  such as that ``every surjection has some section'' (the {\it axiom of choice}), are not regarded as constructively true unless the existence of at least one instance is concretely established. In its modern guise as {\it intuitionistic type theory} (short for: {\it data type} theory!, cf. \cite{Constable11}\cite[\S 5.1]{MySS24}), this paradigm of constructive mathematics means (cf. \cite[p. 42]{MySS24}) that {\it proofs are algorithms} (hence are physical processes, when executed on a mechanical computer): with given assumptions as input, their output constructively witnesses the existence of data of their specified output type. 

Hence if also physical processes are (or were) algorithms, then {\it physical processes are proofs} -- echoing Wittgenstein's identification of ``the world'' with ``all facts''.

\smallskip

\noindent {\bf Generalizing sets to types.}
But to make sense of this, we highlight another lesson of type theory -- famous among specialists (cf. \cite[\S 2.1]{Constable11}) but otherwise underappreciated: As the name indicates, type theory is a foundation of mathematics whose fundamental elementary objects are not necessarily just {\it sets} of isolated {\it elements}. Instead, types:

\begin{itemize}[
  topsep=2pt,
  itemsep=2pt
]
\item[(i)]\hypertarget{ExtraStructure}{}
may carry extra {\it structure} (cf. \cite[p 53]{MySS24}),

\item[(ii)]\hypertarget{PointFree}{} need not be determined by its elements (now called {\it terms}), and 

\item[(iii)]\hypertarget{HigherSymmetry}{} have their (higher categorical) symmetries built-in (cf. \cite[pp 40]{MySS24}).
\end{itemize}
Another way to say this is: Where set theory is realized (only) by the ordinary category of sets, intuitionistic (homotopy) type theory is  realized (``modeled'' by ``semantics'', cf. \cite{Jacobs98}) more generally by categories called (higher) {\it toposes} --- from $\tau\mbox{\'o}\pi{o}\varsigma$ for ``place'':
already according to \cite{Lawvere97}, (cohesive) toposes are where {\it physics may take place} (exposition in \cite{HTTinPhysics25}, more details in \cite{SmoothSets}).

\medskip

\noindent {\bf Space is a cohesive type.}
\begin{itemize}[
  leftmargin=.5cm]
    \item 

As an example for \hyperlink{ExtraStructure}{(i)}: In (homotopy) {\it cohesive type theories} \cite{CoHoTT14}\cite{Shulman18}\cite{Corfield20} to be realized in (higher) {\it cohesive toposes} \cite[\S 3.1]{SS20-Orb}, the real line,  hence {\bf the continuum}
$$
  \mathbb{R}
$$
as understood not just in classical physics but notably in quantum physics (where $\mathbb{C} \simeq \mathbb{R} \times \mathrm{i}\mathbb{R}$), exists, including its smooth- and ring-structure, on the same fundamental level as any plain set. 
% (The trick is that while $\mathbb{R}$ is not itself finite, it can be characterized by a finite number of axioms.) 
This suggests that the notorious trouble that set-based approaches to algorithmic physics have with ``the continuum limit'' may be an artifact of not considering non-discrete cohesive types.

\item As an example for \hyperlink{PointFree}{(ii)}: In {\it super-cohesive} toposes \cite[\S 3.1.3]{SS20-Orb} also the {\bf super-point} 
$$
  \mathbb{R}^{0\vert 1}
$$ 
(having a single element/term $0$, but equipped with a ``fermionic infinitesimal halo'', cf. \cite[Fig. 4]{Freed99}\cite[\S 3.1]{Cherubini18}) exists on the same fundamental level as any set --- in fact including its (abelian) super-Lie algebra structure.

\item As an example for \hyperlink{HigherSymmetry}{(iii)}: The homotopy type of the circle, namely the {\bf classifying space} of the integers (having a single element, but equipped with $\mathbb{Z}$-symmetry): 
$$
  \shape S^1 \,\simeq\,\mathbf{B}\mathbb{Z}
$$
exists on the same fundamental level as plain sets (cf. \cite[(147)]{MySS24}), as do all its ``higher deloopings'' $\mathbf{B}^n\mathbb{Z}$ (having a single element, but equipped with $n$-categorical higher $\mathbb{Z}$-symmetry, cf. \cite[(192)]{MySS24}).

\smallskip 
\item As a combined example: Every homotopy Lie algebra ($L_\infty$-algebra) exists (cf. \cite[\S 4.5.1]{dcct}) as a cohesive homotopy type with a single element but equipped with any infinitesimal higher symmetry.
In particular, for every classifying space $\mathcal{A}$ as in \eqref{NonabelianCohomology}, there exists its {\it Whitehead $L_\infty$-algebra} (\cite[Prop. 5.11]{FSS23-Char})
$$
  \mathfrak{l}\mathcal{A}
$$
such that flat $\mathfrak{l}\mathcal{A}$-valued differential forms 
\cite[Def. 6.1]{FSS23-Char}
are precisely 
flux densities for which $\mathcal{A}$ is an admissible flux-quantization law, as in \S\ref{QuantumGravity} (cf. \cite[\S 3]{SS25-Flux}).
\end{itemize}

\medskip
\noindent {\bf Super-spacetime emerges.}
Like a mustard seed, the super-point $\mathbb{R}^{0\vert 1}$ is tiny and yet carries seminal internal structure. Homotopy types detect this inner structure via a non-trivial 2-cocycle, namely a non-null map of super-$L_\infty$ algebras
\begin{equation}
  \label{2CocycleOnSuperpoint}
  \begin{tikzcd}
    \mathbb{R}^{0\vert 1}
    \ar[
      rr,
      "{
        \mathrm{d}\theta
        \wedge
        \mathrm{d}\theta
      }"
    ]
    &&
    \mathfrak{l}
    \mathbf{B}^2 \mathbb{Z}
    \mathrlap{\,.}
  \end{tikzcd}
\end{equation}
An equivalent incarnation of cocycles are the {\it extensions} which they classify, which in turn are equivalently the {\it homotopy fibers} (cf. \cite[Def. 1.14]{FSS23-Char}\cite[p. 41]{MySS24}) of their classifying map. But for \eqref{2CocycleOnSuperpoint} this turns out to be \cite[p. 18]{Superpoint}
the real {\it super-line}, (or ``super-continuum'')
$$
  \begin{tikzcd}[
    row sep=16pt
  ]
    \mathbb{R}^{1|\mathbf{1}}
    \ar[
      d,
      ->>,
      "{
        \mathrm{hofib}
      }"{swap, pos=.4}
    ]
    \\
    \mathbb{R}^{0\vert\mathbf{1}}
    \ar[
      rr,
      "{
        \mathrm{d}\theta \wedge
        \mathrm{d}\theta
      }"
    ]
    &&
    \mathfrak{l}B^2 \mathbb{Z}
  \end{tikzcd}
$$
equipped with its super-translation structure, hence the $D=1$, $\mathcal{N}=1$ super-symmetry algebra.

Yet more remarkably, the doubled superpoint
$$
  \mathbb{R}^{0\,\vert\,\mathbf{1}\oplus\mathbf{1}}
  \;\simeq\;
  \mathbb{R}^{0\,\vert\,1}
  \underset{
    \mathbb{R}^0
  }{\sqcup}
  \mathbb{R}^{0\,\vert\,1}
$$
carries 3 independent 2-cocycles whose corresponding extension is \cite[Prop. 9]{Superpoint} nothing but $D=3$, $\mathcal{N}=1$ super-spacetime

\vspace{-.6cm}
$$
  \begin{tikzcd}[
    row sep=16pt
  ]
    \mathbb{R}^{
      1,2\,\vert\,\mathbf{2}
    }
    \ar[
      d,
      ->>,
      "{
        \mathrm{hofib}
      }"{swap, pos=.4}
    ]
    \\
    \mathbb{R}^{0\,\vert\,\mathbf{1}\oplus\mathbf{1}}
    \ar[
      rr,
      "{
        \mathbf{d}\theta^{(i}
        \wedge
        \mathbf{d}\theta^{j)}
      }"
    ]
    &&
    \mathfrak{l}B^2 \mathbb{Z}^3
  \end{tikzcd}
$$
with its metric structure encoded in its external automorphism algebra \cite[Prop. 6]{Superpoint}.

Proceeding in this manner by doubling the fermions on this super-space,
its maximal $\mathrm{Spin}(1,2)$-equivariant extension next is
\cite[Thm. 14]{Superpoint}
nothing but $D=4$, $\mathcal{N}=1$ super-spacetime
$$
  \begin{tikzcd}[
    row sep=16pt
  ]
    \mathbb{R}^{
      1,3\,\vert\, \mathbf{4}
    }
    \ar[
      d,
      ->>,
      "{
        \mathrm{hofib}
      }"{swap, pos=.4}
    ]
    \\
    \mathbb{R}^{
      1,2
        \,\vert\,
      \mathbf{2}\oplus \mathbf{2}
    }
    \ar[
      rr
    ]
    &&
    \mathfrak{l}
    B^2 \mathbb{Z}
  \end{tikzcd}
$$
again with its metric structure encoded in its external automorphisms.

This progression continues \cite[Thm. 14]{Superpoint} and discovers next $D=6$, then $D=10$, and finally $D=11$ super-spacetime, see \hyperlink{BraneBouquet}{Figure 2}.
We have hence a kind of {\it emergence of spacetime} from pure computational logic (``It from Bit''), rather different from traditional set-based approaches and right away recovering the continuum structure of spacetime together with its local (super-)metric structure.
\footnote{
More precisely, what emerges here are the Kleinian local model spaces of higher-dimensional supergravities; but from these, 
curved supergravity follows as the super-Cartan geometric extension, cf. \cite{GSS24-SuGra}.
}

\medskip

\noindent {\bf Super-branes emerge.} 
When seen in the higher super-cohesive topos, these super-spacetimes sprout a whole bouquet of further invariant {\it higher extensions} \cite{FSS19-HigherRational} which may be understood \cite{CdAIPB00} as (higher super-spacetimes extended by charges of) brane species: The {\it brane bouquet} \cite{FSS15-WZW}\cite[p. 14]{SS19-ADEEquivariant} shown in \hyperlink{BraneBouquet}{Figure 2}.
Notably, 11D super-space carries an invariant 4-cocycle which is the WZW term of the M2-brane sigma model, whence the higher central extension it classifies is known as $\mathfrak{m}2\mathfrak{brane}$ (\cite[\S 3.1.3]{JH12}\cite[Def. 4.2]{FSS15-WZW}\cite[p. 13]{SS19-ADEEquivariant} to be read as: ``11D super-space extended by M2-brane charges''):
$$
  \begin{tikzcd}[
    column sep=60pt
  ]
    \mathfrak{m}2\mathfrak{brane}
    \ar[
      d,
      ->>,
      "{
        \mathrm{hofib}
      }"{swap, pos=.4}
    ]
    \\
    \mathbb{R}^{
      1,10\,\vert\, 
      \mathbf{32}
    }
    \ar[
      rr,
      "{
        G_4^0
        \,:=\,
        \tfrac{1}{2}
        (\,
          \overline{\psi}
          \,\Gamma_{a_1 a_2}\,
          \psi
        )
        e^{a_1}
        e^{a_2}
      }"
    ]
    &&
    \mathfrak{l}
    B^4\mathbb{Z}\,.
  \end{tikzcd}
$$
This, in turn, carries yet one more invariant 7-cocycle, being the WZW term of the M5-brane sigma model:
$$
  \begin{tikzcd}[
    column sep=85pt
  ]
    \mathfrak{m}5\mathfrak{brane}
    \ar[
      d,
      ->>,
      "{
        \mathrm{hofib}
      }"{swap, pos=.4}
    ]
    \\
    \mathfrak{m}2\mathfrak{brane}
    \ar[
      rr,
      "{
        \widetilde G_7
        \,:=\,
        \tfrac{1}{5!}
        (\,
          \overline{\psi}
          \,\Gamma_{a_1 \cdots a_5}\,
          \psi
        )
        e^{a_1}
        \cdots
        e^{a_5}
        -
        \tfrac{1}{2}
        c_3 \, G_4
      }"
    ]
    &&
    \mathfrak{l}
    B^7\mathbb{Z}\,.
  \end{tikzcd}
$$

\newpage

\begin{tabular}{p{14cm}}
\hypertarget{BraneBouquet}{}
\footnotesize
{\bf Figure 2 -- The Brane Bouquet.}  In cohesive homotopy theory there emerges, from the super-point, a bouquet of (invariant central) {\it higher extensions} which first \cite{Superpoint} grows the super-spacetimes in the critical dimensions of string theory, then
\cite{FSS15-WZW}
sprouts the corresponding super-brane species, and eventually blossoms into the M-brane species on 11D super-space classified in rational 4-Cohomotopy \cite{FSS15-M5WZW}\cite{FSS17} 
(animated exposition in \cite{AnimatedBraneBouquet}, more background in \cite[Fig. 1]{FSS19-HigherRational}\cite[Fig. 3]{SS19-ADEEquivariant}).

This  is a ``structural'' or ``synthetic'' emergence of spacetime, which is possible in topos/type theory, quite distinct in nature from attempts to see spacetime emerge via point-set or graph models.
\end{tabular}

\vspace{.1cm}

\begin{tikzpicture}
\node at (0,0) {
\begin{tikzcd}[decoration=snake]
  &[-12pt]
  &[-24pt]
  &[-15pt]
  \mathfrak{m}5\mathfrak{brane}
  \ar[
    d,
    ->>
  ]
  &[-16pt]
  &[-20pt]
  &[-12pt]
  \\
  &
  &
  &
  \mathfrak{m}2\mathfrak{brane}
  \ar[
    dd,
    ->>
  ]
  \\
  \mathfrak{d}5\mathfrak{brane}
  \ar[
    ddr,
    ->>
  ]
  &
  \mathfrak{d}3\mathfrak{brane}
  \ar[
    dd,
    ->>
  ]
  &
  \mathfrak{d}1\mathfrak{brane}
  \ar[
    ddl,
    ->>
  ]
  &&
  \mathfrak{d}0\mathfrak{brane}
  \ar[
    ddr,
    ->>
  ]
  \ar[
    dl,
    dotted
  ]
  &
  \mathfrak{d}2\mathfrak{brane}
  \ar[
    dd,
    ->>
  ]
  &
  \mathfrak{d}4\mathfrak{brane}
  \ar[
    ddl,
    ->>
  ]
  \\
  \mathfrak{d}7\mathfrak{brane}
  \ar[
    dr,
    ->>
  ]
  &
  &&
  \mathbb{R}^{1,10\,\vert\,\mathbf{32}}
  \ar[
    ddr,
    ->>
  ]
  &&&
  \mathfrak{d}6\mathfrak{brane}
  \ar[
    dl,
    ->>
  ]
  \\
  \mathfrak{d}9\mathfrak{brane}
  \ar[
    r,
    ->>
  ]
  &
  \mathfrak{string}_{\mathrm{IIB}}
  \ar[
    dr,
    ->>
  ]
  &&&&
  \mathfrak{string}_{\mathrm{IIA}}
  \ar[
    dl,
    ->>
  ]
  &
  \mathfrak{d}8\mathfrak{brane}\
  \ar[
    l,
    ->>
  ]
  \\
  &&
  \mathbb{R}^{1,9\,\vert\,\mathbf{16} \oplus \mathbf{16}}
  &
  \mathbb{R}^{1,9\,\vert\,\mathbf{16}}
  \ar[
   r,
   shift left=3.7pt
  ]
  \ar[
   r,
   shift right=3.7pt
  ]
  \ar[
   l,
   shift left=3.7pt
  ]
  \ar[
   l,
   shift right=3.7pt
  ]
  \ar[
    dr,
    ->>
  ]
  &
  \mathbb{R}^{1,9\,\vert\,\mathbf{16} \oplus \overline{\mathbf{16}}}
  \\
  &&&
  \mathbb{R}^{1,5\,\vert\,\mathbf{8}}
  \ar[
    dl,
    ->>
  ]
  \ar[
   r,
   shift left=3.7pt
  ]
  \ar[
   r,
   shift right=3.7pt
  ]
  &
  \mathbb{R}^{1,5\,\vert\,\mathbf{8}\oplus\overline{\mathbf{8}}}
  \\
  &&
  \mathbb{R}^{1,3\,\vert\,\mathbf{4}\oplus\mathbf{4}}
  &
  \mathbb{R}^{1,3\,\vert\,\mathbf{4}}
  \ar[
    dl,
    ->>
  ]
  \ar[
   l,
   shift left=3.7pt
  ]
  \ar[
   l,
   shift right=3.7pt
  ]
  \\
  &&
  \mathbb{R}^{1,2\,\vert\,\mathbf{2}\oplus\mathbf{2}}
  &
  \mathbb{R}^{1,2\,\vert\,\mathbf{2}}
  \ar[
    dl, 
    ->>
  ]
  \ar[
   l,
   shift left=3.7pt
  ]
  \ar[
   l,
   shift right=3.7pt
  ]
  \\
  &&
  \mathbb{R}^{0\,\vert\,\mathbf{1}\oplus\mathbf{1}}
  &
  \mathbb{R}^{0\,\vert\,\mathbf{1}}
  \ar[
   l,
   shift left=3.7pt,
   gray
  ]
  \ar[
   l,
   shift right=3.7pt
  ]
  \ar[
    uuu,
    -Latex,
    shift right=33pt,
    shorten=13pt,
    gray,
    "{
      \scalebox{.7}{
        \color{darkgreen}
        \bf
        \def\arraystretch{.8}
        \begin{tabular}{c}
          (higher)
          \\
          invariant
          \\
          super-Lie
          \\
          extensions
        \end{tabular}
      }
    }"{swap, xshift=-8pt}
  ]
\end{tikzcd}
};
\draw[
  draw opacity=0,
  fill=olive,
  fill opacity=.3
]
 (-1.1,6.3) rectangle 
 (+.8,-6.4);
\node
  at (.7,-5.9) {
      \adjustbox{
      raise=2pt,
      scale=.7
    }{
      \color{darkblue}
      \bf
      \def\arraycolsep{0pt}
      \def\arraystretch{.86}
      \begin{tabular}{c}
        super-
        \\
        point
      \end{tabular}
    }
  };
\node
  at (1.3,2.1) {
      \adjustbox{
      raise=2pt,
      scale=.7
    }{
      \color{darkblue}
      \bf
      \def\arraycolsep{0pt}
      \def\arraystretch{.86}
      \begin{tabular}{c}
        11D super-
        \\
        spacetime
      \end{tabular}
    }
  };
\node
  at (1.1,5.2) {
      \adjustbox{
      raise=2pt,
      scale=.7
    }{
      \color{darkblue}
      \bf
      \def\arraycolsep{0pt}
      \def\arraystretch{.86}
      \begin{tabular}{c}
        super
        \\
        M-branes
      \end{tabular}
    }
  };
\end{tikzpicture}

\medskip

\noindent
{\bf The C-field emerges.}
Finally, the abelian M2- and M5-brane cocycles unify \cite[\S 3]{FSS15-M5WZW}\cite[(57)]{FSS19-HigherRational} into a single non-abelian cocycle in 4-Cohomotopy \cite[Cor. 2.3]{FSS17}, via {\it homotopy pullback} (cf. \cite[Ex. 1.12]{FSS23-Char}) 
along the Hopf fibration \eqref{QuaternionicHopfFibration}:

\vspace{-.55cm}
$$
  \hspace{1.7cm}
  \begin{tikzcd}[
    row sep=17pt
  ]
    \mathfrak{m}5\mathfrak{brane}
    \ar[
      d,
      "{\ }"{name=t1}
    ]
    \ar[
      rr,
      "{\ }"{name=s1, swap, pos=.2}
    ]
    \ar[
      from=s1,
      to=t1,
      Rightarrow,
      "{
        \scalebox{.6}{
          \color{gray}
          (hpb)
        }
      }"{xshift=-4pt}
    ]
    &&
    \ast
    \ar[d]
    \\
    \mathfrak{m}2\mathfrak{brane}
    \ar[
      rr,
      "{ 
        \widetilde G_7 
      }",
      "{\ }"{swap, name=s2, pos=.2}
    ]
    \ar[
      d,
      "{\ }"{name=t2}
    ]
    \ar[
      from=s2,
      to=t2,
      Rightarrow,
      "{
        \scalebox{.6}{
          \color{gray}
          (hpb)
        }
      }"{xshift=-4pt}
    ]
    &&
    \mathfrak{l}S^7
    \ar[
      d,
      "{
        \mathfrak{l}
        h_{\mathbb{H}}
      }"
    ]
    \\
    \mathbb{R}^{
      1,10\,\vert\, \mathbf{32}
    }
    \ar[
      rr,
      "{
        (G_4,\, G_7)
      }"
    ]
    &&
    \mathfrak{l}S^4
  \end{tikzcd}
$$
But this 4-Cohomotopy cocycle in 11D --- which thus emerges from the super-point --- is the avatar (in a precise sense, \cite[Thm. 3.1]{GSS24-SuGra}) of the C-field that we started the discussion with!, above in \eqref{CharacterIn4Cohomotopy}.

\smallskip

This may be seen to close a grand circle, where (super-)gravitational spacetime emerges from homotopical logic (\S\ref{Vista}), as such holographically exhibits topological qbits (\S\ref{QuantumGravity}), which in turn are naturally described in homotopy-typed language (\S\ref{andComputation}).

\end{document}